# Massive Multiple Input Massive Multiple Output for 5G Wireless Backhauling

D.-T. Phan-Huy[1], P. Ratajczak[1], R. D'Errico[2], J. Järveläinen[3*], D. Kong[4], K. Haneda[3], B. Bulut[4], A. Karttunen[3], M. Beach[4], E. Mellios[4], M. Castañeda[5], M. Hunukumbure[6], T. Svensson[7]
[1]Orange Labs, [2]CEA-LETI, [3]Aalto Univ.,*Premix Oy, [4]Univ. of Bristol, [5]Huawei, [6]Samsung, [7]Chalmers Univ. of Technology
dinhthuy.phanhuyorange.com

*Abstract*— In this paper, we propose a new technique for the future fifth generation (5G) cellular network wireless backhauling. We show that hundreds of bits per second per Hertz (bits/s/Hz) of spectral efficiency can be attained at a high carrier frequency (such as 26 GHz) between large antenna arrays deployed along structures (such as lamp posts) that are close and roughly parallel to each other. Hundreds of data streams are spatially multiplexed through a short range and line of sight "massive multiple input massive multiple output" (MMIMMO) propagation channel thanks to a new low complexity spatial multiplexing scheme, called "block discrete Fourier transform based spatial multiplexing with maximum ratio transmission" (B-DFT-SM-MRT). Its performance in real and existing environments is assessed using accurate ray-tracing tools and antenna models. In the best simulated scenario, 1.6 kbits/s/Hz of spectral efficiency is attained, corresponding to 80% of Singular Value Decomposition performance, with a transmitter and a receiver that are 200 and 10,000 times less complex, respectively.

## I. Introduction

Due to the availability of large spectrum at higher carrier frequencies, the spectrum bands corresponding to millimeter waves are good candidates for the self-backhauling of the future 5th generation (5G) of mobile networks [1]. Their coverage limitation could be overcome through a dense deployment. In this paper, we propose to boost the spectral efficiency of millimeter wave based backhaul links through a new type of deployment. In theory [2-5], two uniform linear arrays (ULAs) of $N$ antenna elements and of equal length and parallel to each other, communicating through a line of sight (LOS) multiple input multiple output (MIMO) propagation channel (as illustrated in Fig. 1), can multiplex $N$ data streams in the spatial domain, under the following conditions:

$$\frac{L^2}{\lambda D} = N; \quad (1)$$
$$D \gg L, \quad (2)$$

where $\lambda = c/f$ is the wavelength ($c$ and $f$ being the speed of light and the carrier frequency, respectively), $D$ is the distance between the ULAs, $L$ is the ULAs' length (as illustrated in Fig. 1.). According to [5], conditions (1) and (2) guarantee that the MIMO channel matrix has $N$ equal eigenvalues. Recently, [6] has shown that $N$ can reach values as high as several hundreds of antenna elements, if $\lambda$ corresponds to 5G (candidate) high carrier frequencies and if $D$ and $L$ are chosen smartly. These new types of deployments, that we call "massive multiple input massive multiple output" (MMIMMO), could deliver gigantic spectral efficiencies of hundreds of bits/s/Hz. In this paper, for the first time, we evaluate MMIMMO deployments in real and existing environments, using ray-tracing tools that accurately model the scattering. Antenna radiation patterns are also accurately modeled and practical deployment considerations not exactly fulfilling (1) are also taken into account.

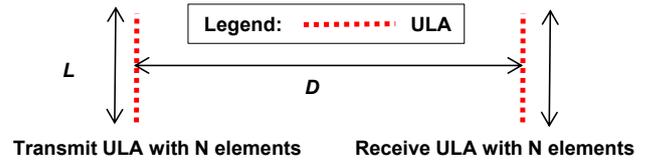

Fig. 1. Communication between two ULAs in LOS.

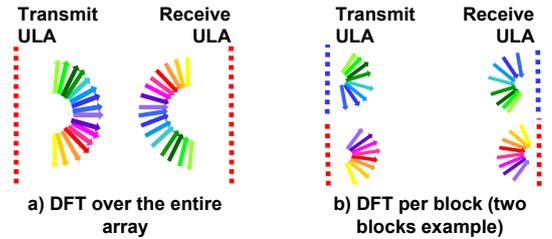

Fig. 2. 16×16 MIMO system mapping 16 streams into 16 angles.

In this paper, we also propose a new practical signal processing scheme for these new MMIMMO deployments. As applying singular value decomposition (SVD) to a MMIMMO system is too complex, we propose to re-use a practical low complexity spatial multiplexing (SM) scheme that combines discrete Fourier transform (DFT) and maximum ratio transmission (MRT) precoding, called DFT-SM-MRT [7]. Fig. 2-a) illustrates the use of DFT-SM alone (without MRT) with two ULAs parallel to each other and in LOS. In this case, data streams are mapped into angles. In DFT-SM-MRT, the role of MRT is to mitigate the effect of scattering and to deal with cases where the ULAs might not be perfectly parallel. However, DFT-SM-MRT [7] still suffers from residual interference, especially when condition (2) is not met. In this paper, we present a new low complexity scheme called block DFT-SM-MRT (B-DFT-SM-MRT), with a similar complexity as the one of DFT-SM-MRT. As illustrated in Fig. 2, it applies the DFT per block. The main idea of this scheme is to approximately fulfill condition (2) on a per-block basis.

The outline of the paper is as follows. Section II defines a set of practical MMIMMO links that could be deployed in existing environments, and their corresponding antenna and propagation models. Section III presents the novel B-DFT-SM-

MRT scheme and recalls the definitions of the DFT-SM-MRT and SVD schemes. Section IV compares these schemes in terms of performance and complexity, for all links defined in Section II. Section V concludes this paper.

The following notations are used. $\mathbf{I}^{(N)}$ is the identity matrix of size $N$. $\mathbf{M}^{\text{DFT}(N)}, \mathbf{M}^{\text{IDFT}(N)} \in \mathbb{C}^{N \times N}$ are the Butler matrices of size $N$, for the DFT and the IDFT operations, respectively. If $\mathbf{A} \in \mathbb{C}^{N \times M}$, $\mathbf{A}^*$ is the conjugate of $\mathbf{A}$, $\mathbf{A}^\dagger$ is the transpose conjugate of $\mathbf{A}$, rank($\mathbf{A}$) is the rank of $\mathbf{A}$, $\mathbf{A}_{n,m}$ is the element in the $n$-th row and $m$-th column, with $1 \leq n \leq N$ and $1 \leq m \leq M$. If $x \in \mathbb{C}$, then $|x|$ is the module $x$.

## II. MMIMMO LINKS IN EXISTING ENVIRONMENTS

To assess the performance of MMIMMO links in real and existing environments, we have built environment-specific channel models. We consider $f = 26 \cdot 10^9$ Hz, as it corresponds to a candidate carrier frequency for 5G in Europe. We consider a narrowband signal. Such signal can either be obtained with a narrowband single carrier waveform or a narrow sub-band of a wideband multi-carrier waveform. With this assumption, the propagation between the transmit array and the receive array can be considered as frequency flat and can be modeled with a complex channel matrix. Let $\mathbf{H} \in \mathbb{C}^{N \times N}$ be the MMIMMO propagation channel matrix. We model the propagation between the $n$-th transmit antenna element and the $m$-th receive antenna with a finite number of rays $N_{n,m}^{\text{rays}}$ that depends on $n$ and $m$. Indeed, different pairs of receive and transmit antenna, which are very far apart in the arrays, may see different numbers of scatterers. The $r$-th ray (with $1 \leq r \leq N_{n,m}^{\text{rays}}$) has a path gain $\alpha_{n,m,r} \in \mathbb{C}$, a direction of arrival vector $\overrightarrow{\mathbf{DoA}_{n,m,r}} \in \mathbb{R}^3$ and a direction of departure vector $\overrightarrow{\mathbf{DoD}_{n,m,r}} \in \mathbb{R}^3$. We assume that all transmit and receive antenna elements have the same antenna gain function $\Gamma \in \mathbb{R}^{\mathbb{R}^3}$, $\Gamma$ being a function of the direction of arrival (or departure). With these notations, the channel coefficient $\mathbf{H}_{n,m}$ between the receive antenna $n$ and the transmit antenna $m$ is given by:

$$\mathbf{H}_{n,m} = \sum_{r=1}^{N_{n,m}^{\text{rays}}} \alpha_{n,m,r} \Gamma(\overrightarrow{\mathbf{DoA}_{n,m,r}}) \Gamma(\overrightarrow{\mathbf{DoD}_{n,m,r}}). \quad (3)$$

We have used two different ray-tracing tools to obtain the parameters $N_{n,m}^{\text{rays}}$, $\alpha_{n,m,r}$, $\overrightarrow{\mathbf{AoA}_{n,m,r}}$ and $\overrightarrow{\mathbf{AoD}_{n,m,r}}$ in two different environments: an outdoor and an indoor environment, described in the Sections II-A and II-B, respectively. Each tool is modelling a real and existing environment in which antenna elements can be positioned. The aforementioned parameters are then generated based on the chosen positions of the antenna elements. $\Gamma(\overrightarrow{\mathbf{DoA}_{n,m,r}})$ and $\Gamma(\overrightarrow{\mathbf{DoD}_{n,m,r}})$ are determined based on an accurate antenna model presented in the Section II-C. Finally, the method for the setting of the MMIMMO parameters (such as the number of antenna elements $N$) is given in Section II-D.

### A. MMIMMO links in the City Center of Bristol

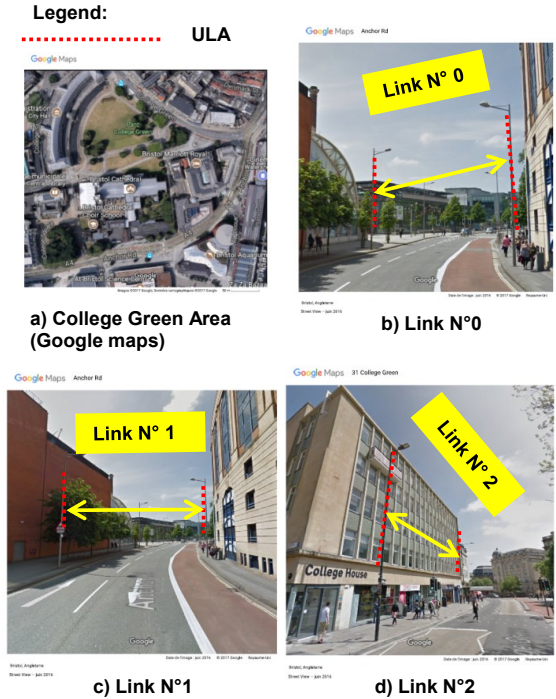

Fig. 3. Modeled links for outdoor environment (in Bristol's City Center).

Fig. 3 a) illustrates the considered outdoor environment. The chosen outdoor environment is an existing road of the "College Green Area", in the city center of Bristol, in the United Kingdom. The ULAs are assumed to be deployed on existing lamp-posts. Three different links between lamp-posts, named "Link N° 0", "Link N° 1" and "Link N° 2" are considered and illustrated in Fig. 3 b), c) and d), respectively. The employed ray tracer identifies the radio wave scatterers using an accurate geometrical database of the physical environment [8], [9]. A similar scenario has been adopted in [10]. Point-source three dimensional (3D) ray-tracing is performed from each antenna element of the transmit array to each antenna element of the receive array assuming isotropic elements. The tool provides the necessary information to compute the parameters of equation (3) for each ray.

### B. MMIMMO links in the Helsinki Airport

As an example of indoor environment, we chose to model the existing Helsinki airport check-in hall. Again, the used ray-tracing tool uses an accurate geometrical database of the physical environment, a so called point cloud model [11]. The point cloud model includes small objects (e.g. self-check-in machines) which scatter energy at high carrier frequencies. Our simulator is calibrated with experimental measurements made in the Helsinki airport check-in hall [12].

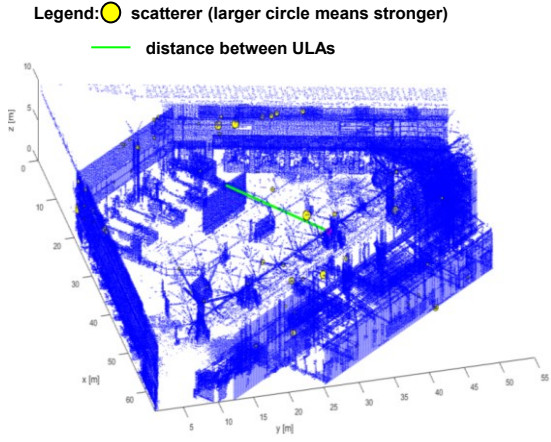

Fig. 4. Point cloud model of the Helsinki Airport Check-In Hall, illustrating the propagation between one point on the giant screen and one point on the canopy: the main LOS direction is indicated by the light green straight line, the scatterers identified by the tool are indicated by the yellow circles. Larger yellow circle means that the scatterer has a stronger impact.

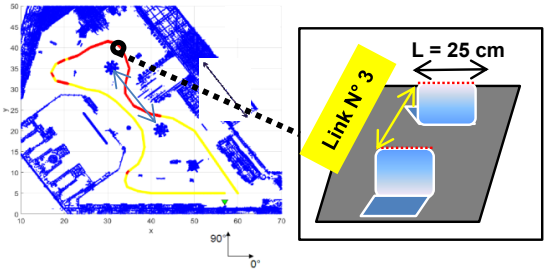

**a) Link N° 3 between laptops in Helsinki Airport Check-In Hall Airport (view of the hall from above)**

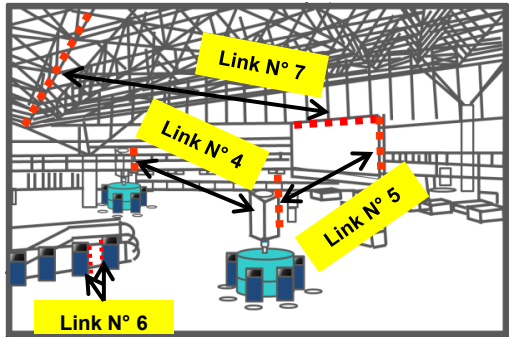

**b) Links N° 4, 5, 6 and 7 in Helsinki Airport Check-In Hall**

Fig. 5. Modeled links for indoor environment in Helsinki Airport Check-In Hall.

As illustrated in Fig. 4, our simulator accurately identifies the locations and the reflection (or scattering) coefficients of the scatterers. The information on the scatterers allows us to derive the parameters of equation (3). Different MMIMMO links illustrated in Fig. 5 are considered: between nearby devices ("Link N° 3"); between signboards ("Link N° 4" and "Link N° 5"); between self-check-in-machines ("Link N° 6") and finally between a signboard and a canopy ("Link N° 7").

## C. Model of antennas

The radiation patterns at 26 GHz of two different antennas are generated by simulation: a 'basic antenna' (a classical printed dipole on a ground plane) illustrated in Fig. 6-a) and a "directional antenna" illustrated in Fig. 6-b). The directional antenna consists of five units of the basic antenna, separated by 1.5 wavelengths. As a finite number of discrete spatial samples of these radiation patterns are generated by simulations, an interpolation between samples is necessary to obtain the exact values of $\Gamma(\overrightarrow{\mathbf{DoA}_{n,m,r}})$ and $\Gamma(\overrightarrow{\mathbf{DoD}_{n,m,r}})$ used in Equation (3).

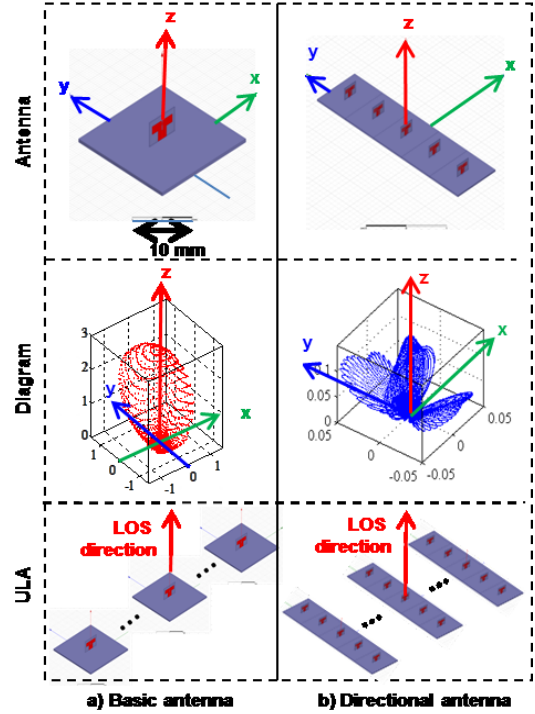

Fig. 6. Antenna radiation pattern of an antenna element and ULAs.

## D. MMIMMO parameters

For each MMIMMO link listed in Section II-A and Section II-B, we compute the ULAs parameters depending on the physical structures (lamp posts, signboards etc.) on which the ULAs are deployed.

Let $N^U$ and $d$ be the number of data streams multiplexed in the spatial domain and the inter-antenna spacing, respectively. The number of antenna elements $N$ is set equal to $N^U$ in the case of the DFT-SM-MRT scheme and the SVD scheme. As it will later be explained in Section III, $N$ potentially very slightly exceeds $N^U$ in the case of B-DFT-SM-MRT. Although, from the ray-tracing tools, we know the exact value of $D$, we compute the MMIMMO system parameters based on an approximation $\widehat{D}$ (with an arbitrarily small chosen error in the order of a decimeter), since we assume that in a real deployment situation, one can only obtain an imperfect measurement of $D$.

We choose $d$ and $N^U$, where $N^U = N$ (for all cases except some configurations of B-DFT-SM-MRT), so that condition (1) is met as much as possible and with the following additional constraint: $N^U$ must be a power of 2. Note that this constraint only applies to $N^U$ and does not apply to $N$ when $N > N^U$. This latter requirement ensures a low complexity implementation of the DFT. Two different methods to chose $N^U$ are tested in this paper.

In the first method (applied to the outdoor links), we arbitrarily set $N^U = 64 = 2^6$. In the second method (applied to the indoor links), we determine the length $L$ (in meters) of the physical structure on which we deploy the ULA. We then compute the largest $N^U$ that is deployable within $L$ and that is close to fulfilling condition (1), as follows:

$$N^U = 2^K \text{ and} \quad (4)$$
$$K = \arg\left\{\max_{k \in \mathbb{N}} \left(2^k \leq \frac{L^2}{\lambda D}\right)\right\}.$$

In practice, one cannot position antennas with an infinite precision. We thus define $\delta$ as the spatial step for the positioning of antennas. $d$ is then determined as follows, for both methods:

$$d = \frac{\delta}{N^U}\left\lfloor\frac{1}{\delta}\sqrt{\lambda D N^U}\right\rfloor.$$

Table I lists the parameters of the considered MMIMMO links. Note that for some links, condition (2) is not met (i.e. $\widehat{D}/(dN_D)$ is not much higher than 1). Compared to DFT-SM-MRT, B-DFT-SM-MRT is therefore expected to improve these links. For the Link N°4, we allow the deployment of antennas 30 cm above the check-in machine height.

TABLE I. MMIMMO PARAMETERS WITH $\delta = 0.1\ mm$

| Link N° | $\widehat{D}$ (m) | $L$ (m) | $N^U$ | $d$ (mm) | $N^U d$ (m) | $\frac{\widehat{D}}{dN^D}$ |
|---|---|---|---|---|---|---|
| 3 | 0.5 | 0.30 | 8 | 26.8 | 0.2144 | ~2 |
| 4 | 17.4 | 1.55+0.30 | 16 | 112 | 1.792 | ~10 |
| 5 | 8.9 | 3.7 |  | 45.8 | 2.9312 | ~3 |
| 0, 1 and 2 | 25 | NA | 64 | 67.1 | 4.2944 | ~6 |
| 6 | 0.9 | 1.3 | 128 | 9 | 1.152 | ~13 |
| 7 | 18.7 | 8.85 | 256 | 28.7 | 7.3472 | ~3 |

III. STUDIED SCHEMES

This section describes the three following spatial multiplexing schemes: A) DFT-SM-MRT [7] (as the baseline method); B) B-DFT-SM-MRT (as the new proposed method); C) SVD spatial multiplexing (as an upper bound). To make a fair comparison, we impose the following common constraint: the number of streams $N_U$ and the inter-antenna spacing $d$ defined in Section II are common to all schemes. As illustrated in Fig. 7, only the following parameters can be scheme-specific: the number of antenna elements $N$, the spatial precoder and the spatial decoder. As a consequence, the propagation channel matrix $\mathbf{H} \in \mathbb{C}^{N \times N}$ and the equivalent channel matrix $\mathbf{G} \in \mathbb{C}^{N_U \times N_U}$ (that includes precoding, propagation, and decoding) are also scheme-specific.

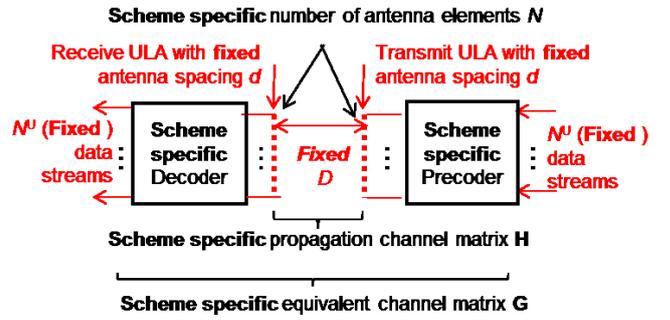

Fig. 7. Common and scheme specific parameters.

To assess the maximum achievable performance of the MMIMMO system, we assume that the signal to noise ratio is very large, and that the system is only limited by the signal to interference ratio (SIR). The SIR of one data stream can be derived based on the MIMO equivalent channel matrix $\mathbf{G}$. The SIR $\mathbf{sir}_n$ of each data stream $n$ is given by:

$$\mathbf{sir}_n = \frac{|\mathbf{G}_{n,n}|^2}{\sum_{p=1, p \neq n}^{N^U}|\mathbf{G}_{n,p}|^2}, 1 \leq n \leq N^U.$$

The theoretical attainable spectral efficiency $\mathbf{c}_n$ for the data stream number $n$ is given by: $\mathbf{c}_n = \log_2(1 + \mathbf{sir}_n), 1 \leq n \leq N^U$. Practical modulations (such as 256 QAM or QPSK) and coding schemes have a bounded spectral efficiency. We therefore define the minimum and maximum spectral efficiencies, $s^{MIN}$ and $s^{MAX}$, accordingly. We define the practical spectral efficiency $\mathbf{c}_n^P$ as follows: $\mathbf{c}_n^P = \min(\mathbf{c}_n, s^{MAX})$ if $\min(\mathbf{c}_n, s^{MAX}) > s^{MIN}$ and $\mathbf{c}_n^P = 0$ otherwise. The resulting total spectral efficiency $s$ is therefore:

$$s = \sum_{p=1}^{N^U} \mathbf{c}_n^P. \quad (5)$$

For each spatial multiplexing scheme, the spectral efficiency is determined using equation (5), this equation being fed with a scheme-specific expression of $\mathbf{G}$.

Next sub-sections provide the expressions of the scheme-specific parameters $\mathbf{H}$, $\mathbf{G}$ and $N$. The same transmit power constraint is assumed for all transmitters.

A. *DFT-SM-MRT*

For DFT SM-MRT [7], the number of antenna elements is $N = N^U$. As illustrated in Fig. 8, $\rho \mathbf{H}^\dagger \mathbf{M}^{IDFT(N^U)}$ is the precoder and $\mathbf{M}^{DFT(N^U)}$ is the decoder, with $\rho$ being a scheme-specific normalising factor to satisfy the power constraint. The equivalent MIMO channel $\mathbf{G}$ is thus:

$$\mathbf{G} = \rho \mathbf{M}^{DFT(N^U)} \mathbf{H} \mathbf{H}^\dagger \mathbf{M}^{IDFT(N^U)}. \quad (6)$$

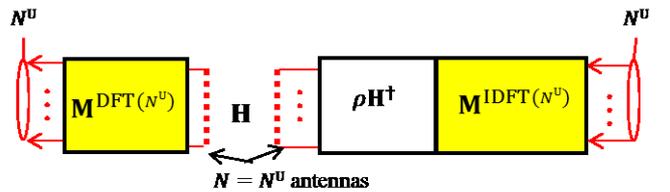

Fig. 8. DFT-SM-MRT spatial multiplexing scheme [7].

## B. B-DFT-SM-MRT

As illustrated in Fig. 9, compared to DFT-SM-MRT, B-DFT-SM-MRT applies the DFT to $N^S$ blocks of $N^D$ data symbols separately, with $N^D = N^U/N^S$. $N^S$ is selected so that condition (2) is better fulfilled, at least on a per-block basis, i.e. such that: $\widehat{D}/(dN^D) > 1$. We optionally append a cyclic prefix (CP) [13] of $N^{CP}$ symbols in the spatial domain (with $0 \leq N^{CP} \leq N^D$), after each 'per block' DFT operation. As symbols are mapped onto antennas, this has a direct impact on the role of each antenna. The $N^S$ blocks of $N^D$ data symbols are mapped onto $N^S$ blocks of $N^D$ 'data antennas'. This constitutes a set of $N^U$ 'useful antennas'. $N^S$ blocks of $N^{CP}$ symbols are mapped onto $N^S$ blocks of $N^{CP}$ 'CP antennas'. Each block of $N^{CP}$ 'CP antennas' is inserted between two successive blocks of $N_D$ 'data antennas'. Each block of $N^D$ data streams goes through an inverse DFT, which is equivalent to a multiplication by $\mathbf{M}^{\text{IDFT}(N^D)}$. We set $N^E = N^D + N^{CP}$. This time, $N \geq N^U$ antenna elements (instead of $N^U$) are used at both the transmitter and receiver sides, with:

$$N = N^S(N^D + N^{CP}) = N^S N^E = N^U + N^S N^{CP}. \quad (7)$$

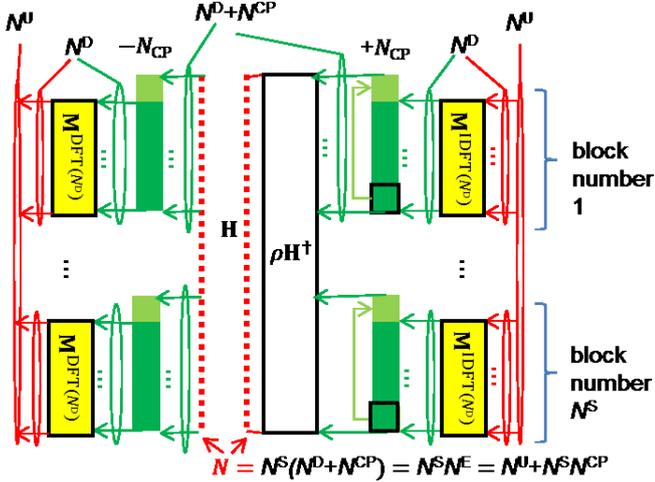

Fig. 9. B-DFT-SM-MRT spatial ultiplexing scheme.

We define the matrices $\mathbf{A}, \mathbf{A}' \in \mathbb{C}^{N^E \times N^D}$, $\mathbf{B}, \mathbf{B}' \in \mathbb{C}^{N^D \times}$, $\mathbf{T} \in \mathbb{C}^{N \times N^U}$ and $\mathbf{R} \in \mathbb{C}^{N^U \times N}$ as follows:

- $\mathbf{A}_{k,l} = 1$ if $l = N^D + k$ and $1 \leq k \leq N^{CP}$ or if $l = k$ and $N^{CP} + 1 \leq k \leq N^{CP} + N^D$; $\mathbf{A}(k,l) = 0$ otherwise;
- $\mathbf{B}_{k,l} = 1$ if $l = N^{CP} + k$ and $1 \leq k \leq N^D$; $\mathbf{B}(k,l) = 0$ otherwise;
- $\mathbf{A}' = \mathbf{A}\mathbf{M}^{\text{IDFT}(N^D)}$ and $\mathbf{B}' = \mathbf{M}^{\text{DFT}(N^D)}\mathbf{B}$;
- $\mathbf{T}_{k+(n-1)N^E, l+(n-1)N^D} = \mathbf{A}'_{k,l}$, for $1 \leq n \leq N^S$, $1 \leq k \leq N^E$ and $1 \leq l \leq N^D$;
- $\mathbf{R}_{k+(n-1)N^D, l+(n-1)N^E} = \mathbf{B}'_{k,l}$, for $1 \leq n \leq N^S$, $1 \leq k \leq N^D$ and $1 \leq l \leq N^E$.

With these definitions, the equivalent channel $\mathbf{G}$ is given by:

$$\mathbf{G} = \rho \mathbf{R}\mathbf{H}\mathbf{H}^{\dagger}\mathbf{T}, \quad (8)$$

where $\rho$ is a scheme-specific normalising factor to satisfy the power constraint. Note that when $N^S = 1$ and $N^{CP} = 0$, B-DFT-SM-MRT is identical to DFT-SM-MRT.

## C. SVD

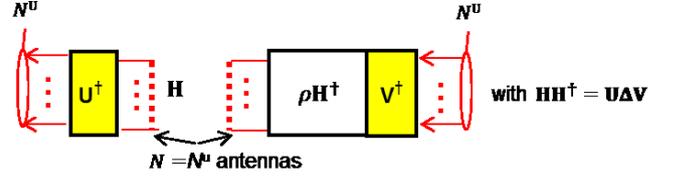

Fig. 10. SVD spatial multiplexing scheme.

The number of antennas $N$ for this scheme is: $N = N^U$. As illustrated in Fig. 10, $\mathbf{U}, \mathbf{V}, \mathbf{\Delta} \in \mathbb{C}^{N^U \times N^U}$ are matrices obtained from the singular value decomposition of $\mathbf{H}\mathbf{H}^{\dagger}$, i.e., such that $\mathbf{H}\mathbf{H}^{\dagger} = \mathbf{U}\mathbf{\Delta}\mathbf{V}$, with $\mathbf{\Delta}$ being diagonal, and $\mathbf{V}\mathbf{V}^{\dagger} = \mathbf{U}\mathbf{U}^{\dagger} = \mathbf{I}^{(N^U)}$. Let $\rho$ be a scheme-specific normalising factor to satisfy the power constraint. $\rho\mathbf{H}\mathbf{V}^{\dagger}$ is the precoder and $\mathbf{U}^{\dagger}$ is the decoder. With these notations, the equivalent MIMO channel $\mathbf{G}$ is:

$$\mathbf{G} = \rho\mathbf{\Delta}. \quad (9)$$

## IV. PERFORMANCE AND COMPLEXITY EVALUATION

The performance analysis and the complexity analysis are performed for the MMIMMO links defined in Section II and the schemes described in Section III. Section IV. A lists the simulated scenarios. Section IV-B and Section IV-C describe the spectral efficiency and complexity evaluation methods, respectively. Finally, Section IV-D provides the results.

## A. Simulation scenarios

Table II lists the simulated scenarios and their corresponding parameters. In this table, and throughout this paper, the notations * and ** indicate that B-DFT-SM-MRT without CP and B-DFT-SM-MRT with CP are used, respectively. The absence of these notations indicates that DFT-SM-MRT is used.

For all scenarios, the channel and antenna models described in Section II are used to generate $\mathbf{H}$.

The performance is also evaluated in a free space (FS) propagation scenario (i.e. a pure LOS scenario). Let $\delta_{n,q}$ be the distance between the receive antenna element $n$ and the transmit antenna element $q$. For FS, $\mathbf{H}_{n,q}$ is given by: $\mathbf{H}_{n,q} = (\frac{\lambda}{4\pi\delta_{n,q}})e^{-j2\pi\delta_{n,q}/\lambda}$.

TABLE II. SIMULATED SCENARIOS

| N° | Link | $N$ | $Nd$ (m) | $N^S$ | $N^D$ | $N^{CP}$ | $\frac{\widehat{D}}{dN^D}$ |
|---|---|---|---|---|---|---|---|
| 3 | 3 | 8 | 0.2144 | 1 | 8 | 0 | ~2 |
| 3* | | | | 2 | 4 | | ~5 |
| 4 | 4 | 16 | 1.792 | 1 | 16 | 0 | ~10 |
| 4* | | | | 2 | 8 | | ~19 |

| | | | | | | |
|---|---|---|---|---|---|---|
| 5 | 5 | | 2.9312 | 1 | 64 | 0 | ~3 |
| **5\*** | | | | 4 | 16 | | ~12 |
| 0 | 0 | 64 | 4.2944 | 1 | 64 | 0 | ~6 |
| **0\*** | | | | 2 | 32 | | ~12 |
| 1 | 1 | | 4.2944 | 1 | 64 | 0 | ~6 |
| **1\*** | | | | 2 | 32 | | ~12 |
| 2 | 2 | | 4.2944 | 1 | 64 | 0 | ~6 |
| **2\*** | | | | 2 | 32 | | ~12 |
| 6 | 6 | 128 | 1.152 | 1 | 128 | 0 | ~1 |
| **6\*** | | | | 16 | 8 | | ~13 |
| **6\*\*** | | 144 | 1.296 | 16 | 8 | 1 | ~13 |
| 7 | 7 | 256 | 7.3472 | 1 | 256 | 0 | ~3 |
| **7\*** | | | | 8 | 32 | | ~20 |
| **7\*\*** | | 264 | 7.5768 | | 32 | 1 | ~20 |

### B. Spectral efficiency evaluation methodology

The spectral efficiency $s$ is computed using Equation (5) and the method given in Section III. We set $s = 8$ bits/s/Hz (corresponding to 256-QAM and a coding rate of 1) and $s^{MIN} = 1$ bit/s/Hz (corresponding to QPSK and a coding rate of 1/2). Note that, for SVD, the spectral efficiency is simply given by $s = N^U s^{MAX}$. For each scheme, the two following metrics are computed:

- the ratio $\phi^{SVD}$ between the spectral efficiency of the considered scheme and the spectral efficiency of SVD;
- the ratio $\phi^{FS}$ between the spectral efficiency of the considered scheme and the spectral efficiency of the same scheme in a FS environment.

The closer to these metrics, the better the schemes are.

### C. Complexity evaluation

We assume a fully digital architecture and we base our complexity evaluation on [14].

We recall that the complexities of the DFT of size $N$, of the SVD of a matrix of size $N \times N$ and of the multiplication of two matrices of sizes $N \times M$ and $M \times P$, scale with $O(N\log_2(N))$, $O(N^3)$ and $O(NMP)$, respectively. As a consequence, $N^S$ DFTs of complexity that scales with $O((N^U/N^S)\log_2(N^U/N^S))$ each, result in a total complexity that scales with $O(N^U \log_2(N^U/N^S))$.

We define $\mathbf{x} \in \mathbb{C}^{N^U \times 1}$ as the vector of transmitted symbols. $\mathbf{z}^{SVD} \in \mathbb{C}^{N^U \times 1}$, $\mathbf{z}^{DFT} \in \mathbb{C}^{N^U \times 1}$ and $\mathbf{z}^{BDFT} \in \mathbb{C}^{N^U + N^S N^{CP}}$ are the vectors of symbols received at the receive antenna array for the SVD, the DFT-SM-MRT and the B-DFT-SM-MRT, respectively. Using these notations, we derive the complexities scaling laws for the transmitter (taking into account the spatial precoding only) and the receiver (taking into account the spatial decoding) and report them in Table III. Our analysis excludes the MRT block as it appears in all the compared schemes (SVD included).

We finally define $\mu^{TX}$ (and $\mu^{RX}$, respectively) as the ratio of the complexity scaling law of the transmitter (the receiver respectively) of SVD, over the complexity scaling law of the transmitter (the receiver respectively) of the considered scheme. Using the expressions in Table III, we obtain the expressions of $\mu^{TX}$ and $\mu^{RX}$, for the DFT-SM-MRT and the B-DFT-SM-MRT, in Table IV. The larger these metrics, the better they are. Indeed, the transmitter (respectively the receiver) of the considered scheme, is $\mu^{TX}$ (respectively $\mu^{RX}$) less times complex than the one of SVD.

TABLE III. SVD SCHEME COMPLEXITY SCALING LAWS (TX= TRANSMITTER, RX= RECEIVER)

| | | Computations | Complexity scaling law |
|---|---|---|---|
| SVD | Tx | $\mathbf{V}, \mathbf{V}^\dagger \mathbf{x}, \mathbf{H}^\dagger \mathbf{V}^\dagger \mathbf{x}$ | $O\left(N^{U^3} + N^{U^2} + N^{U^2}\right)$ |
| | Rx | $\mathbf{U}, \mathbf{U}^\dagger \mathbf{z}^{SVD}$ | $O\left(N^{U^3} + N^{U^2}\right)$ |
| DFT-SM- | Tx | $\mathbf{M}^{IDFT(N_U)}\mathbf{x}$ $\mathbf{H}^\dagger \mathbf{M}^{IDFT(N_U)}\mathbf{x}$ | $O\left(N^U \log_2(N^U) + N^{U^2}\right)$ |
| | Rx | $\mathbf{M}^{DFT(N_U)}\mathbf{z}^{DFT}$ | $O\left(N^U \log_2(N^U)\right)$ |
| B-DFT-SM- | Tx | $\mathbf{H}^\dagger \mathbf{T}.\mathbf{x}$ | $O\left((N^U + N^S N^{CP})^2 + N^U \log_2\left(\frac{N^U}{N^S}\right)\right)$ |
| | Rx | $\mathbf{R}\mathbf{z}^{BDFT}$ | $O\left(N^U \log_2\left(\frac{N^U}{N^S}\right)\right)$ |

TABLE IV. EXPRESSIONS OF $\mu^{TX}$ AND $\mu^{RX}$

| Scheme | $\mu^{TX}$ | $\mu^{RX}$ |
|---|---|---|
| DFT-SM-MRT | $\frac{\log_2(N^U) + N^U}{N^{U^2} + 2N^U}$ | $\frac{\log_2(N^U)}{N^{U^2} + 2N^U}$ |
| B-DFT-SM-MRT | $\frac{(N^U + N^S N^{CP})^2 + N^U \log_2\left(\frac{N^U}{N^S}\right)}{N^{U^3} + 2N^{U^2}}$ | $\frac{\log_2\left(\frac{N^U}{N^S}\right)}{N^{U^2} + N^U}$ |

### D. Simulation results

Table V provides the simulation results for all scenarios listed in Section IV-A). DFT-SM-MRT and B-DFT-SM-MRT both attain spectral efficiencies of several hundreds of bits/s/Hz (that are close to the ones of SVD) with much less complex transmitters and receivers. B-DFT-SM-MRT outperforms DFT-SM-MRT with an even simpler receiver and a slightly more complex transmitter. For all MMIMMO links, except for Link N° 7, the performance is close to the FS performance. This confirms that in the chosen existing environments, the propagation is dominated by LOS and that simple spatial multiplexing schemes (such as DFT-SM-MRT or B-DFT-SM-MRT) can be used. However, for Link N° 7, the performance is much lower than the FS one. The scatterers of scenario 7 are visible on Fig. 4. One can observe that a strong dominating scatterer is located on the metallic ceiling of the canopy. For this particular scenario, we replace the basic antennas by directional antennas (defined in section III-C) oriented along the main LOS direction. We obtain an improved performance that is reported in Table VI. In particular, for scenario 7*, around 1.6 kbits/s/Hz of spectral efficiency is attained, corresponding to 80% of SVD performance with a transmitter and a receiver that are 200 and 10000 less complex, respectively. The CP insertion slightly improves the performance in scenarios 6** and in scenario 7** (with directional antennas). An extensive study of the CP insertion is for further study.

TABLE V. RESULTS († INDICATES DIRECTIONAL ANTENNAS ARE USED)

| N° | SE (bits/s/Hz) | $\phi^{FS}$ (%) | $\phi^{SVD}$ (%) | $\mu^{TX}$ | $\mu^{RX}$ |
|---|---|---|---|---|---|
| 0 | 246 | 65 | 48 | 60 | 693 |
| **0*** | **258** | **62** | **50** | **61** | **832** |
| 1 | 296 | 75 | 58 | 60 | 693 |
| **1B*** | **342** | **81** | **67** | **61** | **832** |
| 2B | 194 | 64 | 38 | 60 | 693 |
| **2B*** | **277** | **75** | **54** | **61** | **832** |
| 3B | 47 | 100 | 74 | NA | NA |
| **3B*** | **52** | **100** | **81** | | |
| 4C | 81 | 63 | 63 | NA | NA |
| **4C*** | **83** | **65** | **65** | | |
| 5D | 349 | 91 | 68 | 60 | 693 |
| **5D*** | **431** | **97** | **84** | **62** | **1040** |
| 6D | 290 | 97 | 57 | 52 | 2359 |
| 6D* | 318 | 99 | 31 | 123 | 5504 |
| **6D**** | **429** | **88** | **42** | **127** | **5504** |
| 7 | 113/**1174†** | 9/**94†** | 6/**57†** | 101 | 8224 |
| 7* | 281/**1651†** | 16/**93†** | 14/**81†** | 253 | 13158 |
| **7**** | **266/1681†** | **15/90†** | **13/82†** | **238** | **13158** |

## V. CONCLUSION

In this paper, we showed that there is an opportunity for future 5G networks operators to exploit the existing urban architecture to transport, on the wireless media, huge data rates with gigantic spectral efficiencies. A new precoding/decoding scheme is proposed, called "block discrete Fourier transform based spatial multiplexing with maximum ratio transmission", B-DFT-SM-MRT, which has a low complexity compared to singular value decomposition. The performance of this scheme at 26 GHz is assessed in existing environments that are accurately modeled with ray-tracing tools. Antennas as well, are accurately modeled. In the best scenario, 1.6 kbits/s/Hz is attained, corresponding to 80% of SVD performance, with a transmitter and a receiver that are 200 and 10000 times less complex, respectively. Further studies will be conducted with measured MMIMMO channel data.

ACKNOWLEDGMENTS

This work has been partially funded by the 5G PPP project mmMAGIC [15] under grant ICT-671650. We warmly thank Mr Antonio Clemente and Mrs Marie-Hélène Hamon for their support on this activity.